\documentclass[10pt,conference, final]{IEEEtran}
\usepackage{cite}
\usepackage{amsmath,amssymb,amsfonts}
\usepackage{algorithmic}
\usepackage{graphicx}
\usepackage{tabularx}
\usepackage{textcomp}
\usepackage{xcolor}
\usepackage{tikz}
\usepackage{url}
\definecolor{Ao}{rgb}{0.0, 0.5, 0.0}
\usepackage[hidelinks]{hyperref}

\pdfoutput=1

\begin{document}
\bstctlcite{IEEEexample:BSTcontrol}

\title{AllTheDocks road safety dataset: \\A cyclist's perspective and experience\\
\thanks{AllTheDocks?, LOTI?, LDW?.}
}

\author{\IEEEauthorblockN{Chia-Yen Chiang\IEEEauthorrefmark{1}, Ruikang Zhong\IEEEauthorrefmark{1}, Jennifer Ding\IEEEauthorrefmark{2}, Joseph Wood\IEEEauthorrefmark{3}, Stephen Bee\IEEEauthorrefmark{4}, Mona Jaber\IEEEauthorrefmark{1}} \IEEEauthorblockA{\IEEEauthorrefmark{1}School of Electronic Engineering and Computer Science\\
Queen Mary University of London, London, UK\\
Email: \{c.chiang, r.zhong, m.jaber\}@qmul.ac.uk} \IEEEauthorblockA{\IEEEauthorrefmark{3}The Alan Turing Institute, British Library, 96 Euston Road, London NW1 2DB,
UK\\
Email: jding@turing.ac.uk}
\IEEEauthorblockA{\IEEEauthorrefmark{3}City, University of London, 
Northampton Square,
London EC1V 0HB, 
UK\\
Email: J.D.Wood@city.ac.uk}
\IEEEauthorblockA{\IEEEauthorrefmark{4}Zwing's, 
UK\\
Email: stephenhbee@outlook.com}
}

\maketitle

\begin{abstract}
Active travel is an essential component in intelligent transportation systems. Cycling, as a form of active travel, shares the road space with motorised traffic which often affects the cyclists' safety and comfort and therefore peoples' propensity to uptake cycling instead of driving.  This paper presents a unique dataset, collected by cyclists across London, that includes video footage, accelerometer, GPS, and gyroscope data. The dataset is then labelled by an independent group of London cyclists to rank the safety level of each frame and to identify objects in the cyclist's field of vision that might affect their experience. Furthermore, in this dataset, the quality of the road is measured by the international roughness index of the surface, which indicates the comfort of cycling on the road. The dataset will be made available for open access in the hope of motivating more research in this area to underpin the requirements for cyclists' safety and comfort and encourage more people to replace vehicle travel with cycling. 
\end{abstract}
\vspace{0.2cm}
\begin{IEEEkeywords}
Cycling, Deep learning, Intelligent transportation, Road safety 
\end{IEEEkeywords}

\section{Introduction}\label{sec:intro}

In order to reduce environmental pollution and excessive carbon emissions caused by urban transportation systems, active travel has become a widespread solution for intra-city travel \cite{winters2017policies}. However, cycling, as a major active travel option, still faces challenges in terms of safety and comfort compared to travelling by vehicle.
The World Health Organisation\footnote{https://www.who.int/news/item/13-12-2023-despite-notable-progress-road-safety-remains-urgent-global-issue} reported an alarming $20\%$ increase in fatal cyclist accidents between 2010 and 2021. A study by the European Commission shows that cyclist deaths in Europe in 2020 represent $10\%$ of all road fatalities with $58\%$ of those occurring in urban environments~\cite{EU}. 

Fortunately, advances in electronics and computer technologies, such as machine learning, computer vision, and inferencing have provided new tools for investigating and identifying opportunities to improve the cycling experience \cite{hagenauer2017comparative}. These emerging technologies can analyse and address the challenges encountered by cyclists from a variety of perspectives. For instance, machine learning approaches summarized in \cite{silva2020machine} can intelligently identify road quality, providing clues for improving road safety. However, it is important to note that the application of these technologies often requires a significant amount of training data \cite{paullada2021data}, which has currently become an impediment to the development of machine learning solutions. Although computer simulations and statistics-based models can also be adapted for analysis and AI model training, datasets collected from the physical world provide information that is difficult to replace \cite{brink2016real}.

This paper is focused on the road safety of cyclists in an urban environment and presents a new dataset collected in London using \textit{citizen science methods} \cite{fraisl2022citizen}. Citizen science initiatives for real-world data collection provide new opportunities for building on-the-ground datasets that are logistically and financially prohibitive otherwise. These methods have been explored for applications such as environmental science, wildlife monitoring, and personal and population health, leveraging low-power, smart sensors to create new, rich datasets for analysis. Within urban studies and analytics, there is an opportunity to apply these techniques to collect new and better datasets on the quality and experience of urban infrastructure and environments.

\href{https://www.londondataweek.org/#about}{London Data Week} is organised by the Greater London Authority (GLA), London Office of Technology and Innovation (LOTI), and The Alan Turing Institute for a weeklong, citywide festival encouraging residents to learn, create, discuss, and explore how to use data to shape our city for the better. For London Data Week 2023, \href{https://allthedocks.com/}{All the Docks} proposed an event centred on a cycling challenge where five teams of three cyclists traverse London to visit all 800+ \href{https://tfl.gov.uk/modes/cycling/santander-cycles}{Santander Cycles} docking stations in one day. In the process, the teams traversed over 250 miles of London roads. As part of the challenge, they decided to collect data along the way on the road and cycling infrastructure that could be shared openly afterwards for road safety research and other public interest analysis.

Cyclists' road safety can be affected by three different factors according to~\cite{SOHAIL2023105949}: road surface conditions, road users' behaviour, and road traffic. A dataset of road accidents involving cyclists is curated in~\cite{Cyclands} based on 30 different datasets taken over a range of 42 years (1979-2021) and 18 cities across three continents. Key features pertaining to the circumstantial conditions of each accident are categorised including road surface conditions, weather, location, and other recorded contextual information. This is a seminal work that sheds light on the relationship between cycling conditions and the risk and severity of accidents. Other works underscore the importance of safety and comfort for increasing the uptake of cycling, in particular, the role of cycle infrastructure, pavement conditions, and road safety conditions. For instance, \cite{mdpi} presents a method for contrasting the observed risk and the user perception of road safety through collected data that includes GPS and front and back video footage. However, the data is not publicly available and is limited to just 30 mins.

Given the aforementioned deficiencies in the comprehensiveness and coverage of existing data on cycling road conditions, we are committed to collecting and publishing a dataset to inform the study of cycling safety and experiences. This work presents and describes the open dataset collected during the All the Docks challenge in July 2023 ad covering $116.15$ kilometers of London roads. The dataset includes visual data and motion-driven sensors' data in addition to GPS information. Understanding the impact of road conditions and furniture on a cyclist's experience and perception of safety is crucial in informing authorities on effective initiatives for encouraging cycling. This work aims to encourage research in this area using open-access data and related code.

We describe the data collection method in Section~\ref{sec:meth} and the dataset and pre-processing methods in Section~\ref{sec:data}. The availability and usage of the data are detailed in Setion~\ref{sec:AandU} and the future work in Section~\ref{sec:next}.


\section{Data collection method and description}\label{sec:meth}

\subsection{Data collection device}\label{sec:dev}
In order to collect a rich dataset of London’s road and cycling infrastructure while in motion, the teams used GoPro cameras as the data collection device as this form of action camera is designed for high-speed, mobile data collection. The GoPros were worn on helmets whereby a single cyclist on a team was equipped with the device to gather image data on the journey, as well as other telemetry data that is captured by GoPros in \href{https://gopro.github.io/gpmf-parser/}{GoPro Metadata Format (GPMF)}. 

\subsection{Routing plans and challenges}\label{sec:route}
In order to visit all the pre-set number of Santander Cycles docking stations, the team utilised a \href{https://oobrien.com/2023/07/all-the-docks-2/} {routing algorithm} to assign each of the five teams to several stations as well as the order they would complete the route. The routing algorithm was devised so teams would be able to start by visiting stations on the periphery of London and end in a central location where all five teams would meet at the end of the challenge. Each route was also designed to be roughly equal in length, though each team faced different challenges along the way such as out-of-service docking stations and stations where there was no available space to dock a bike. During the six-hour-long challenge, some teams also had equipment issues due to the limited power supply of the GoPro. Out of the five teams participating in the All the Docks challenge, two teams had access to a GoPro and backup power supply and were able to collect data on their cycling journey. In future iterations, there is an opportunity to improve the GoPro setup for more consistent, comprehensive data collection.

\subsection{Dataset description}\label{sec:data}

 
There are several types of telemetric datapoints encoded in GroPro video data. These include GPS location, GPS speed, accelerometer, gyroscope, 
magnetometer, gravity vector, and wind detection (see full list in open source software \hyperlink{https://github.com/gopro/gpmf-parser}{gpmf-parser}). \textit{AllTheDocks} dataset is composed of two subsets: video footage and time-lapse photos. The former includes video footage over {$61.68$} kilometres with telemetry data including GPS location, GPS speed, accelerometer, and gyroscope, as detailed in Table~\ref{tab:GoPro}. The latter includes still images sampled at {$0.2$~Hz (every 5 seconds)} and covers {$54.47$} kilometres with only GPS information. 

\begin{table}[]\label{tab:GoPro}
\caption{Telemetry data streams extracted from GoPro video footage.}
\centering
\begin{tabular}{|l|l|l|}
\hline
\textbf{DataType} & \textbf{Units}             & \textbf{Subsets}              \\ \hline
GPS location      & deg°, min', sec"           & Latitude, Longitude \\ \hline
GPS height      & m above/below sea level           & Altitude \\ \hline
GPS speed         & m/s                        & Speed2D, Speed3D                        \\ \hline
Acceleration      & m/s$^2$                    & AcclX, AcclY, AcclZ, Accl3D           \\ \hline
Gyroscope         & rad/s                      & GyroX, GyroY, GyroZ           \\ \hline
\end{tabular}
\end{table}
GPS data including location and speed is sampled around $50$~Hz while the accelerometer and gyroscope data is sampled at $200$~Hz. 
Given this difference in time resolution, we applied linear interpolation to the GPS data to adjust to the higher frequency of $200$~Hz. Next, we concatenated the data from GPS, accelerometer, and gyroscope, now at the same time resolution, and we dropped data points that had empty values.

Given the high sampling frequency of accelerometer data, it is seen to fluctuate much faster than the GPS data. For instance, in the video footage {'GX010049.mp4' on Kingsway road}, both acceleration and GPS speeds are shown in Figure~\ref{fig:enter-label}. It can be seen that there is a correlation between these but that the former signal fluctuates much faster. In Figure~\ref{fig:enter-label}, the peaks of the accelerometer speed signal are marked with a red dot. The highest of those occurs just before $300$~seconds and it can be seen in the video footage that the cyclist experiences a sudden drop due to a pothole. All other marked peaks in~Figure~\ref{fig:enter-label} are caused by rough roads except the one just before $300$~seconds marked by a green triangle which is due to the cyclist changing direction after the sudden drop.  We posit that the accelerometer data captures the behaviour of cyclists which might be affected by road quality, obstacles, or by moving vehicles in their vicinity and would therefore cause them to suddenly steer away. However, our initial analysis shows that it is difficult to infer the cause of such behaviour (e.g., sudden steering) by only examining the accelerometer data. This is majorly due to the GoPro being mounted on the helmets of cyclists thus, the motion sensors were affected by the head movement.
In contrast, accelerometer data can be analysed to inform on the roughness index of a road which impacts the cyclists' comfort, as discussed in Section~\ref{sec:IRI}.

\begin{figure}
    \centering
    \includegraphics[width=0.9\linewidth]{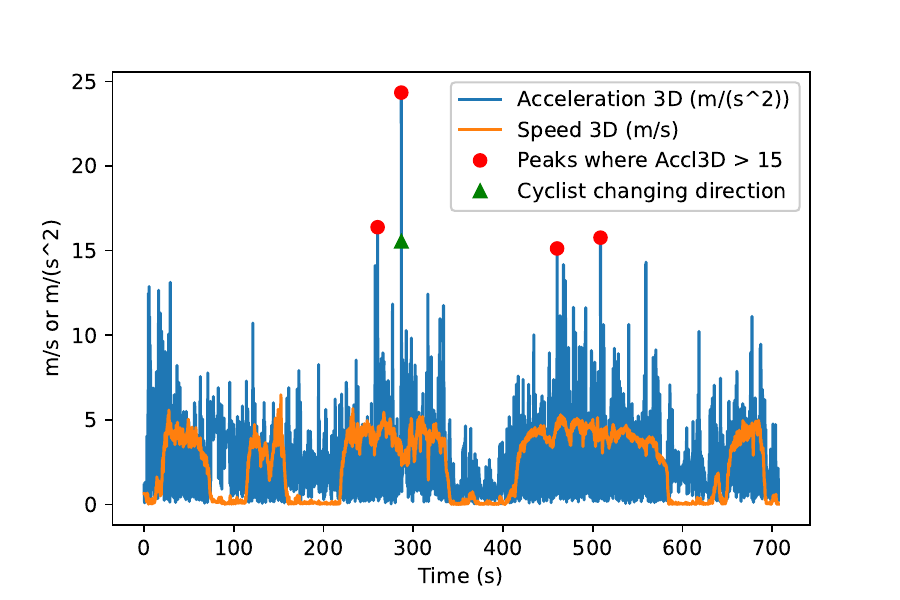}
    \caption{The accelerometer data fluctuates more than GPS speed. The maximum peak of 3D acceleration is caused by a sudden drop on the road, showing the strong influence of road roughness on acceleration.}
    \label{fig:enter-label}
\end{figure}


\begin{figure}
    \centering
    \includegraphics[width=0.8\linewidth]{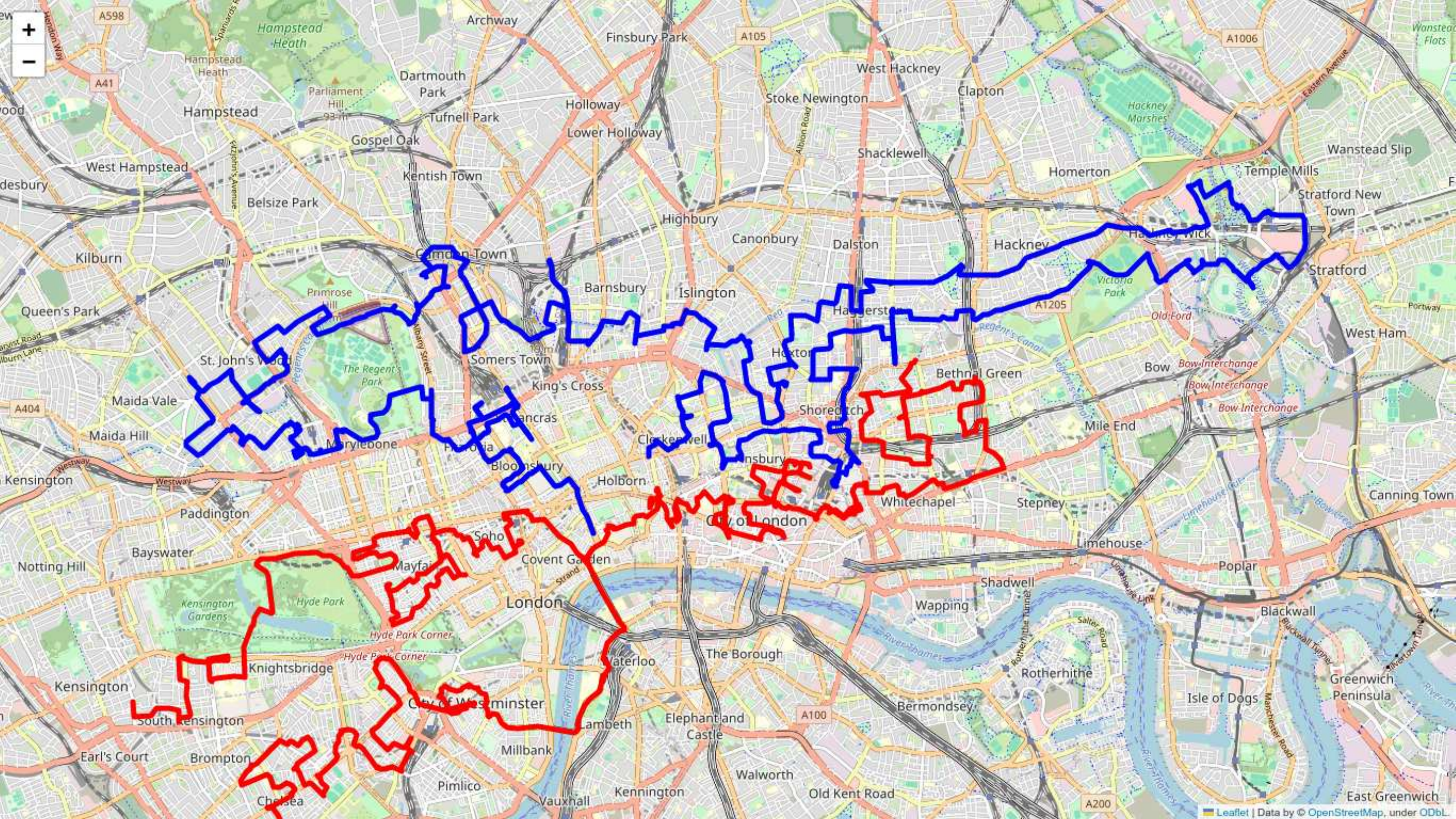}
    \caption{Trajectory covered; (Blue) routes that are represented with video footage and (Red) the route of timelapse images.}
    \label{fig:enter-label}
\end{figure}
\begin{figure}
    \centering
    \includegraphics[width=0.8\linewidth]{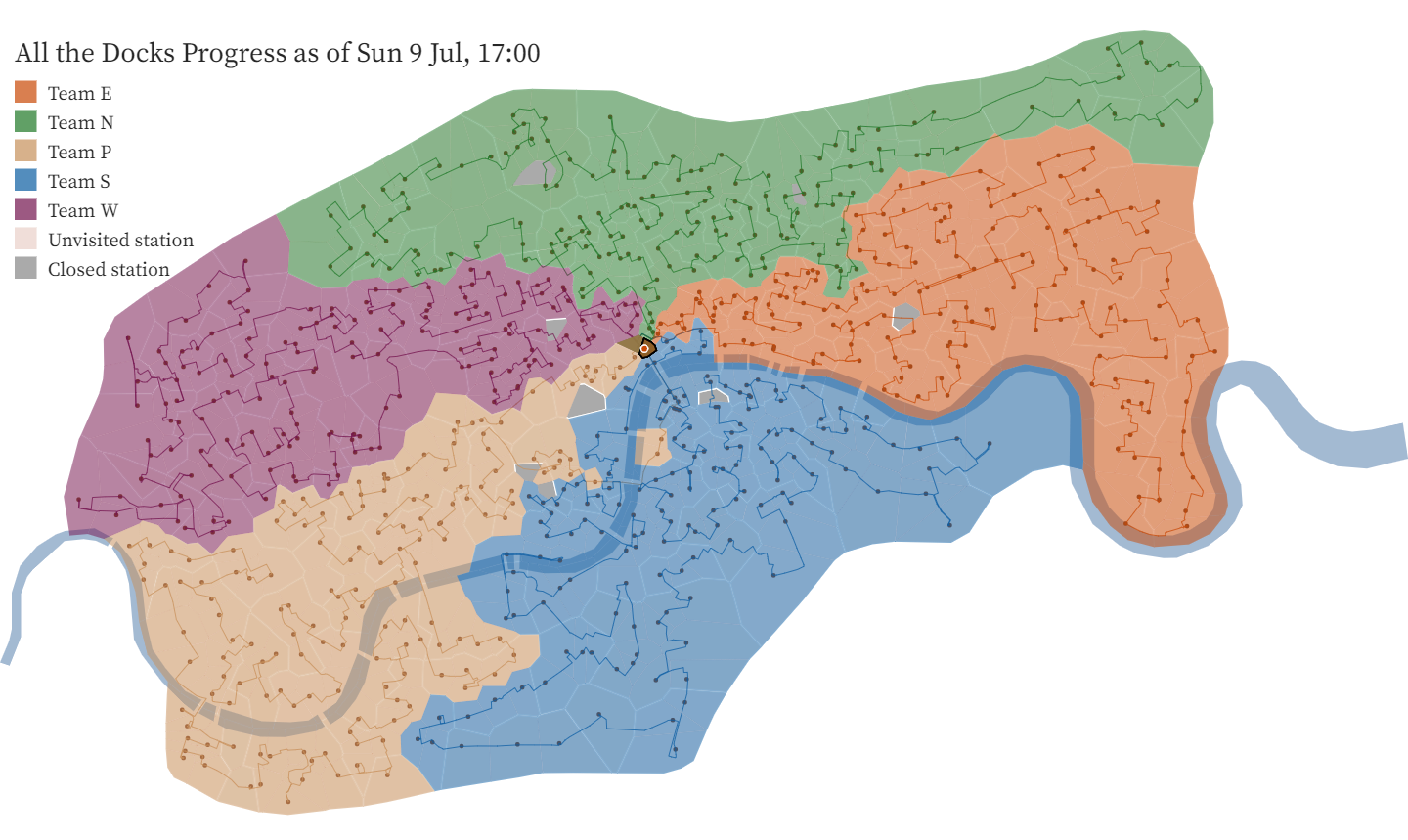}
    \caption{Entire span of routes covered during the All The Docks challenge by 5 teams.}
    \label{fig:enter-label}
\end{figure}
\begin{figure}
    \centering
    \includegraphics[width=0.8\linewidth]{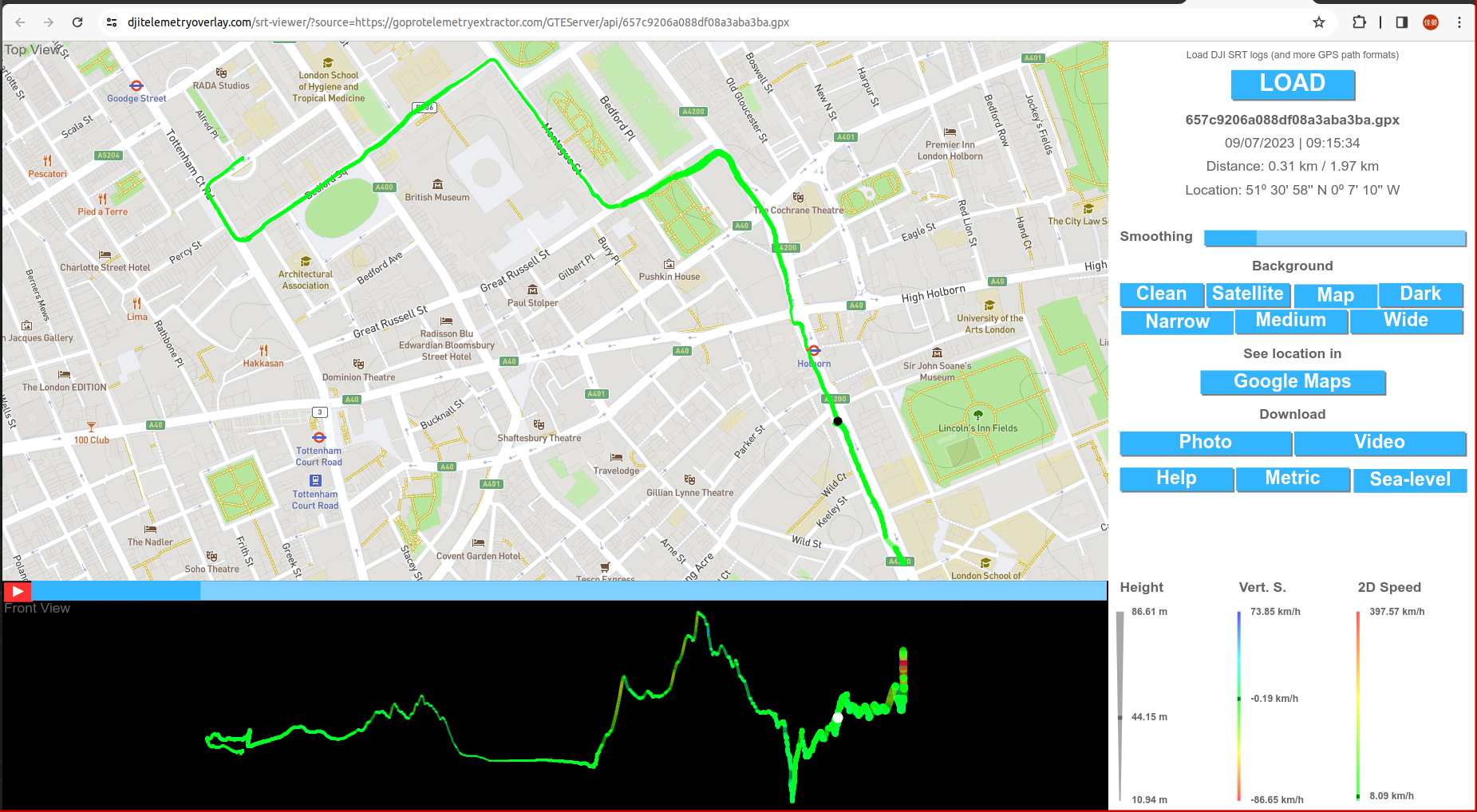}
    \caption{Visualisation of telemetry data embedded in GoPro video footage using \href{https://goprotelemetryextractor.com/free/}{TELEMETRY EXTRACTOR for GoPro}}
    \label{fig:enter-label}
\end{figure}
\section{Road Surface Roughness}\label{sec:IRI}
The surface roughness of roads is an important road characteristic, and it is closely related to the comfort and safety of road users and informative for preemptive road maintenance. With the advent of connected sensors, crowd-sourcing platforms, and machine learning, a prioritisation road repair system is proposed in~\cite{edwin} based on computer vision for detecting potholes and cracks. The road quality for non-motorised traffic is investigated in~\cite{IRI} using GPS and accelerometer data to calculate the International Roughness Index (IRI), which is one of the most widely acknowledged index for measuring road surface roughness.

\subsection{International Roughness Index (IRI)}
IRI is computed based on the vertical displacements over a travelled distance, which is given by:
\begin{equation}
\mathrm{IRI} := \frac{\sum |h_i - h_{i-1}| }{D}.  \label{eq:IRI1}
\end{equation}
 
where $h_i$ is the $i$th vertical displacement in metres, i.e.\ the elevation at time $t_i$, and $D$ is the travelled distance in metres. $AllTheDock$ dataset does not include $h_i$ and $D$ but they can be acquired by processing the raw data from GPS and accelerometer as follows. 

The sum of vertical displacements $\sum |h_i - h_{i-1}|$ can be calculated by a double-integral involving the vertical component of the acceleration $\alpha_s$, where $\mathrm{\alpha_s} = \text{Accl}X_s - g$  at time $s$ and $g=9.81$ meters per second squared is the gravitational acceleration (see Eq.(\ref{eq:ver_dis})). $\text{Accl}X_s$ is the vertical acceleration sampled by the accelerometers, as detailed in table~\ref{tab:GoPro}. The time stamp $s$ is defined such that, $s \in [0, t]$ and $t \in [0, t_{max}]$ where $t_{max}$ is the total time of a cyclist trip.

\begin{equation}
\sum |h_i - h_{i-1}| := \int_0^{t_{\mathrm{max}}} \left|\int_0^t \alpha_s \mathrm{d}s \right| \mathrm{d}t. \label{eq:ver_dis}
\end{equation}








The total travelled distance $D$ can be calculated by integrating the absolute GPS 3D speed $\text{Speed3}D_t$ from time $0$ to time $t_{max}$, as shown in Eq.(\ref{eq:IRI3}). The GPS 3D speed $\text{Speed}3D_t$ is as detailed in table~\ref{tab:GoPro}. The difference between 2D and 3D speed is that the latter includes the speed of the vertical axis (altitude) but not in the former. The 2D speed is a combined speed along longitude and latitude axis.

\begin{equation}
    D = \int_0^{t_{\mathrm{max}}} |\text{Speed}3D_t| \mathrm{d}t \label{eq:IRI3}
\end{equation}

The IRI is a measurement of the quality of a whole section of the road instead of a single spot of the road. Therefore, it gives a single numerical score for the time interval $[0,t_{\mathrm{max}}].$
To see a change over time, we measure the IRI for a series of sliding windows, where each window $\omega_{t_l}$ covers 1000 samples collected over 5 seconds between $t_l$, the starting timestamp of the window and time $t_u=t_l+1000$, the ending timestamp of the window. 
%
In this case, an $IRI_{t_l}$ is computed for every 5 seconds and is associated with the position $(t_u+t_l)/2$, as shown in Figures~\ref{fig:IRI_explain} and~\ref{fig:IRI_thres}.

It is worth noting that the calculated IRI in the \textit{AllTheDocks} dataset is impacted by a few practical uncertainties, as explained in the following paragraph. 

\begin{itemize}
\item Firstly, given that the AllTheDock dataset was collected with the GoPro camera equipped on the cyclist's helmet, the telemetry data is affected by the head movement and may lead to inaccurate IRI calculation (see \textcolor{Ao}{$\blacktriangle$ }in Fig.~\ref{fig:IRI_thres} around $300$ seconds and $650$ seconds). 

\item Secondly, IRI can be increasingly exaggerated by decreasing travelled distance when a cyclist is stopping. The situation generates IRI peaks that are irrelevant to road roughness (see most of the blue peaks in Fig.~\ref{fig:IRI_explain} where the orange curve is flattened; for example, at around $0$, $100$, $200$, $350$, and $600$ seconds).  
\end{itemize}

 \subsection{Modified IRI}
\begin{figure}
    \centering
    \includegraphics[width=0.9\linewidth]{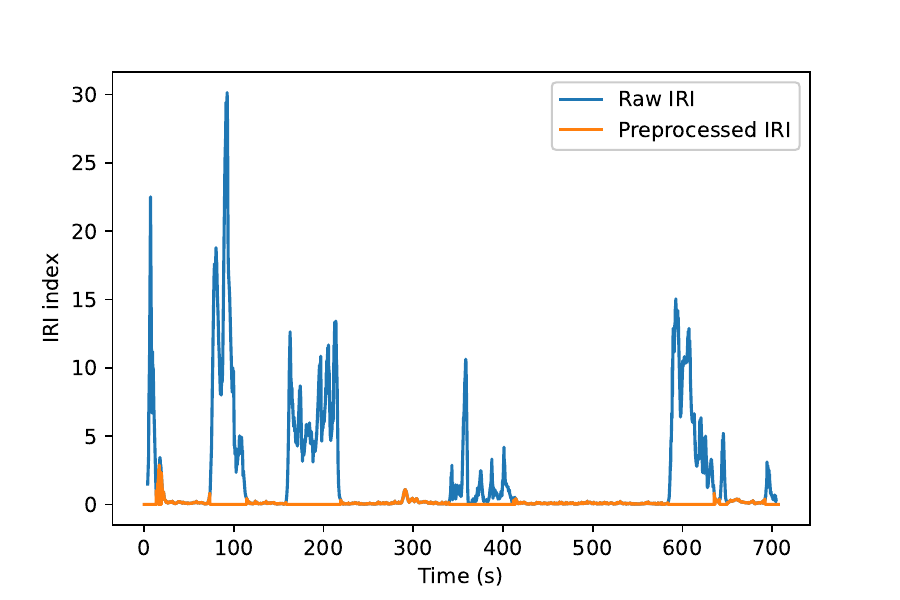}
    \caption{When cyclists are stopping, it generates IRI peaks that are irrelevant to road roughness the same as most peaks in raw IRI (blue) have shown. (Note: in each time s, IRI calculation covers a road section at time s$\pm 2.5$ seconds since our IRI time window is set to 5 seconds.)}
    \label{fig:IRI_explain}
\end{figure}

\begin{figure}
    \centering
    \includegraphics[width=0.9\linewidth]{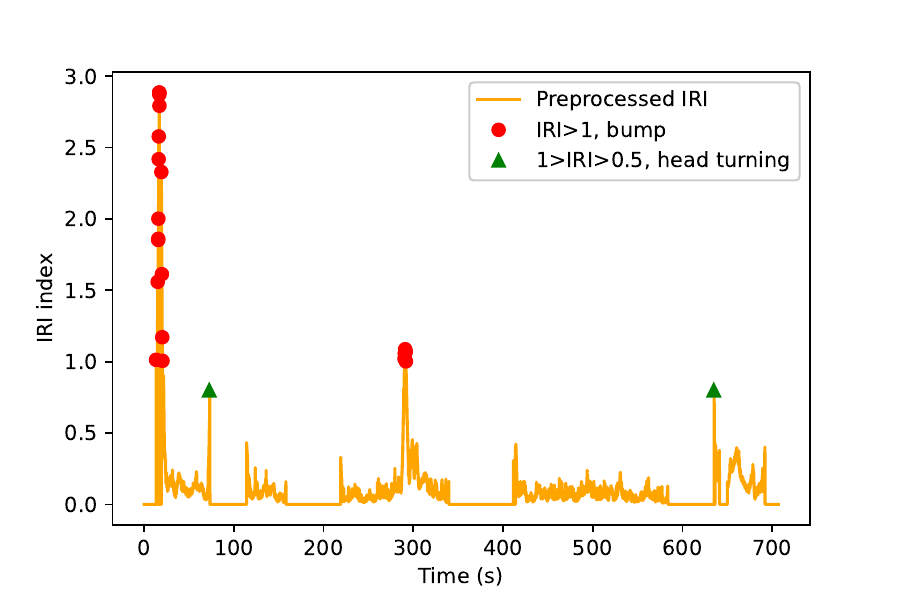}
    \caption{Peak analysis shows it is important to set the IRI threshold properly to remove stopping and head turning moments.}
    \label{fig:IRI_thres}
\end{figure}

To obtain an accurate evaluation of road safety, we need to address the issues mentioned above. Both situations, i.e. head movement and stopping, mostly occur when cyclists are close to a red traffic signal or are docking the bicycle, thus, we filter out these occurrences in the modified IRI. To this end, we force-set the IRI value to zero when the cyclist's 3D speed based on GPS data is less than $1$ m/s (see preprocessed IRI in Fig. \ref{fig:IRI_explain}). IRI peak analysis shows it is important to set the IRI threshold properly to filter out both situations in time-series IRI graph (ig.~\ref{fig:IRI_thres})  

 Finally, time-series IRI can differ when a selected time window changes its size significantly. To intuitively explain, the IRI window is similar to a filter. The wider the window, the more features are detected at each window of time. However, wide windows also smooth out the local/small feature which may lead to small bumps with very small IRI. Therefore, it is necessary to set a proper IRI time window according to features of interest.

\begin{table*}[t]
\centering
\caption{Metadata of GoPro images and videos}
\begin{tabular}{|p{3cm}|p{5cm}|p{2.5cm}|p{3cm}|}
\hline
& Number of files & Available data type & Number of data points for each type\ \\ \hline
GoPro Images (9.2GB)& 3,774 images (Resolution: 4000x3000) GoPro takes a photo every 5 seconds& \begin{enumerate}
    \item  GPS location \item GPS height \end{enumerate}& 3,774 (Time gap: 5 seconds)\\ \hline
GoPro Videos (videos: 117GB, images: 2.9GB)& 32 videos, extracted 4,151 images (Resolution: 1920x1080) We extract a photo from a video every 5~seconds& \begin{enumerate}
    \item GPS location \item GPS height \item GPS speed \item Acceleration \item Gyroscope \item IRI
\end{enumerate}& 4,062,524 (Time gap: 5~milliseconds)\\ \hline
Total   & 32 videos, 7,925 images    & -    & 4,066,298 \\ \hline
\end{tabular}
\label{tab:meta}
\end{table*}
\section{User Perception of Road Safety}\label{sec:label}

\begin{figure}
    \centering
    \includegraphics[width=0.85\linewidth]{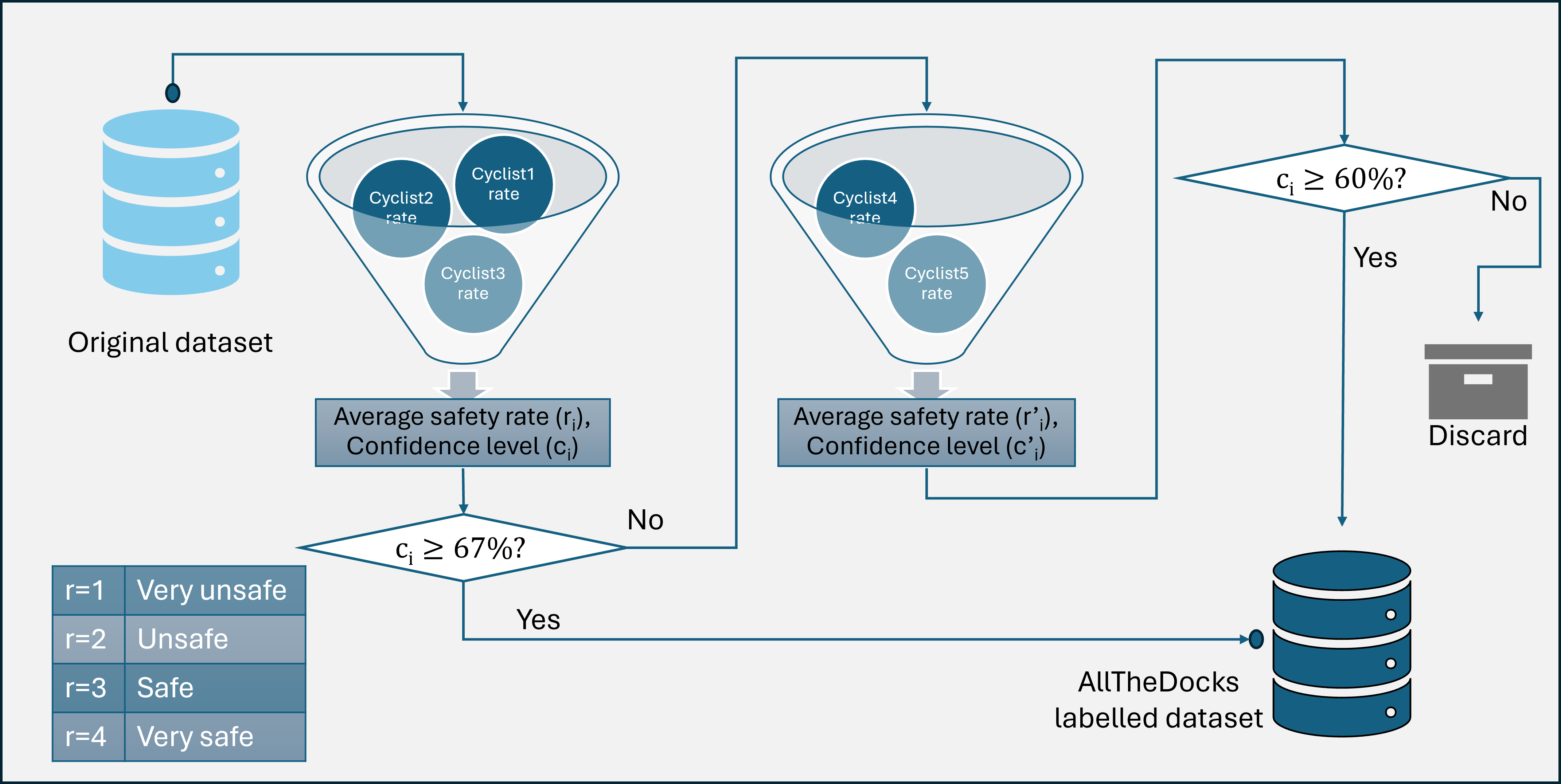}
    \caption{Road safety labelling process}
    \label{fig:label}
\end{figure}

In general, road safety is a subjective matter where the same road and traffic conditions may seem alarming and stressful to one cyclist and very safe to another. However, cyclists who are accustomed to cycling in the same city have common expectations and a more uniform perception of road safety. To this end, we recruited a participant team, where the team members have sufficient cycling experience in London to rate the safety of each frame in the \textit{AllTheDocks} dataset (video-sampled and time-lapse images). Following the approach adopted in~\cite{mdpi}, we employed the 4-point Likert scale to force an opinion from each participant: $[(r=1, Very~unsafe);(r=2, Unsafe);(r=3, Safe);(r=4, Very~safe)]$. As shown in Figure~\ref{fig:label}, each frame $f_i$ is first rated by every cyclist in the participant team, and the average safety rate $r_i$ and confidence level $c_i$ are calculated. If at least half of the cyclists in the team agree on the rating, it is adopted as the safety label, and the frame $f_i$ is added to the final dataset. Otherwise, two additional cyclists are asked to label the frame, and new average safety rate $r'_i$ and confidence level $c'_i$ are calculated. If at least one additional member agrees on the rating, it is adopted and the frame is added to the dataset; otherwise, it is disregarded.
In addition to the safety labelling, we've asked another participant team to identify areas in a frame that might affect the safety or comfort of a cyclist. The items that have been labelled are the following: Pothole, Manhole, Cyclist lane, Pedestrian crossing, and Cycling lane separator.

All labeled data will be released and available online upon the acceptance of the paper. At the moment, we summarise the metadata of GoPro images and videos in the table \ref{tab:meta}.

\section{Availability and usage of data}\label{sec:AandU}
The current plan to open source the \textit{AllTheDocks} dataset is to host and make the dataset available on the \href{https://data.london.gov.uk/}{London Datastore} so that researchers, local government teams, and members of the public can explore and use the data. To do so, we will work with members from the Greater London Authority and Data for London team and address the technical challenges and ethical considerations, following their \href{https://data.london.gov.uk/guidance/sharing-data/}{guidance on sharing data}. Because the London Datastore was created primarily for storing numerical and statistical data, it will be an interesting challenge to work through how to upload an image-based data set. Additionally, we will ensure the data is licensed properly, for example with a Creative Commons license.

\section{Future work}\label{sec:next}
In addition to improving the data collection setup and consistency, the All the Docks challenge has sparked a conversation about new applications that this dataset can help investigate. Beyond cataloging the conditions of road and cycling infrastructure, it is also an opportunity to understand the characteristics of the cycling experience in terms of safety, accessibility, and other quality factors. One of the Greater London Authority partners has also raised the idea of using this data collection method during the nighttime to assess whether cycle infrastructure is designed for night use and how this differs across the city. Because sunlight hours in London are reduced for most of the fall, winter, and early spring, this is an important consideration for the on-the-ground experience of cyclists in the city. Finally, a plan of collecting cycling path data in the winter is also under consideration, as the roads may be icy and slippery in winter, and these rain puddles along the cycling lanes are unique factors that might affect cycling behavior and safety in the winter.

\enlargethispage{-0.489in}
\bibliographystyle{IEEEtran}
\bibliography{IEEEabrv,mybibfile}

\begin{thebibliography}{10}
\providecommand{\url}[1]{#1}
\csname url@samestyle\endcsname
\providecommand{\newblock}{\relax}
\providecommand{\bibinfo}[2]{#2}
\providecommand{\BIBentrySTDinterwordspacing}{\spaceskip=0pt\relax}
\providecommand{\BIBentryALTinterwordstretchfactor}{4}
\providecommand{\BIBentryALTinterwordspacing}{\spaceskip=\fontdimen2\font plus
\BIBentryALTinterwordstretchfactor\fontdimen3\font minus \fontdimen4\font\relax}
\providecommand{\BIBforeignlanguage}[2]{{%
\expandafter\ifx\csname l@#1\endcsname\relax
\typeout{** WARNING: IEEEtran.bst: No hyphenation pattern has been}%
\typeout{** loaded for the language `#1'. Using the pattern for}%
\typeout{** the default language instead.}%
\else
\language=\csname l@#1\endcsname
\fi
#2}}
\providecommand{\BIBdecl}{\relax}
\BIBdecl

\bibitem{winters2017policies}
M.~Winters, R.~Buehler, and T.~G{\"o}tschi, ``Policies to promote active travel: evidence from reviews of the literature,'' \emph{Current environmental health reports}, vol.~4, pp. 278--285, 2017.

\bibitem{EU}
{European Commission}, ``(2023) facts and figures, mobility and amp; transport - road safety.'' European Road Safety Observatory. Brussels, European Commission, Directorate General for Transport, Tech. Rep., 2023.

\bibitem{hagenauer2017comparative}
J.~Hagenauer and M.~Helbich, ``A comparative study of machine learning classifiers for modeling travel mode choice,'' \emph{Expert Systems with Applications}, vol.~78, pp. 273--282, 2017.

\bibitem{silva2020machine}
P.~B. Silva, M.~Andrade, and S.~Ferreira, ``Machine learning applied to road safety modeling: A systematic literature review,'' \emph{Journal of traffic and transportation engineering (English edition)}, vol.~7, no.~6, pp. 775--790, 2020.

\bibitem{paullada2021data}
A.~Paullada, I.~D. Raji, E.~M. Bender, E.~Denton, and A.~Hanna, ``Data and its (dis) contents: A survey of dataset development and use in machine learning research,'' \emph{Patterns}, vol.~2, no.~11, 2021.

\bibitem{brink2016real}
H.~Brink, J.~Richards, and M.~Fetherolf, \emph{Real-world machine learning}.\hskip 1em plus 0.5em minus 0.4em\relax Simon and Schuster, 2016.

\bibitem{fraisl2022citizen}
D.~Fraisl \emph{et~al.}, ``Citizen science in environmental and ecological sciences,'' \emph{Nature Reviews Methods Primers}, vol.~2, no.~1, p.~64, 2022.

\bibitem{SOHAIL2023105949}
\BIBentryALTinterwordspacing
A.~Sohail, M.~A. Cheema, M.~E. Ali, A.~N. Toosi, and H.~A. Rakha, ``Data-driven approaches for road safety: A comprehensive systematic literature review,'' \emph{Safety Science}, vol. 158, p. 105949, 2023. [Online]. Available: \url{https://www.sciencedirect.com/science/article/pii/S0925753522002880}
\BIBentrySTDinterwordspacing

\bibitem{Cyclands}
\BIBentryALTinterwordspacing
M.~Costa, M.~Marques, C.~Roque, and F.~Moura, ``Cyclands: Cycling geo-located accidents, their details and severities,'' \emph{Sci Data}, vol.~9, p. 237, 2022. [Online]. Available: \url{https://www.nature.com/articles/s41597-022-01333-2}
\BIBentrySTDinterwordspacing

\bibitem{mdpi}
\BIBentryALTinterwordspacing
S.~Cafiso, G.~Pappalardo, and N.~Stamatiadis, ``Observed risk and user perception of road infrastructure safety assessment for cycling mobility,'' \emph{Infrastructures}, vol.~6, no.~11, 2021. [Online]. Available: \url{https://www.mdpi.com/2412-3811/6/11/154}
\BIBentrySTDinterwordspacing

\bibitem{edwin}
\BIBentryALTinterwordspacing
E.~Salcedo, M.~Jaber, and J.~Requena~Carrión, ``A novel road maintenance prioritisation system based on computer vision and crowdsourced reporting,'' \emph{Journal of Sensor and Actuator Networks}, vol.~11, no.~1, 2022. [Online]. Available: \url{https://www.mdpi.com/2224-2708/11/1/15}
\BIBentrySTDinterwordspacing

\bibitem{IRI}
\BIBentryALTinterwordspacing
K.~Zang, J.~Shen, H.~Huang, M.~Wan, and J.~Shi, ``Assessing and mapping of road surface roughness based on gps and accelerometer sensors on bicycle-mounted smartphones,'' \emph{Sensors}, vol.~18, no.~3, 2018. [Online]. Available: \url{https://www.mdpi.com/1424-8220/18/3/914}
\BIBentrySTDinterwordspacing

\end{thebibliography}

\end{document}